\documentclass[conference]{IEEEtran}
\usepackage{amsmath,graphicx}
\usepackage{comment}
\usepackage{setspace}
\usepackage{epstopdf}
\usepackage{algorithm}
\usepackage{algorithmic}
\usepackage{tikz}
\usetikzlibrary{trees}
\usepackage{multirow}
\usepackage{caption}

\def\D{{\cal D}}
\def\Q{{\cal Q}}
\def\F{{\cal F}}

\ifCLASSINFOpdf
\else
\fi
\begin{document}
\title{ShakeMe: Key Generation From Shared Motion}


%\author{\IEEEauthorblockN{H{\i}d{\i}r Y\"{u}z\"{u}g\"{u}zel}
%\IEEEauthorblockA{Department of Signal Processing \\ Tampere University of Technology\\ Tampere, Finland\\
%Email: hidir.yuzuguzel@tut.fi}
%\and
%\IEEEauthorblockN{Jari Niemi}
%\IEEEauthorblockA{Department of Signal Processing \\ Tampere University of Technology\\ Tampere, Finland\\
%Email: jari.a.niemi@tut.fi}
%\and
%\IEEEauthorblockN{James Kirk\\ and Montgomery Scott}
%\IEEEauthorblockA{Starfleet Academy\\
%San Francisco, California 96678-2391\\


\author{\IEEEauthorblockN{H{\i}d{\i}r Y\"{u}z\"{u}g\"{u}zel\IEEEauthorrefmark{1},
Jari Niemi\IEEEauthorrefmark{1},
Serkan Kiranyaz\IEEEauthorrefmark{2}, 
Moncef Gabbouj\IEEEauthorrefmark{1} and
Thomas Heinz\IEEEauthorrefmark{3}}
\IEEEauthorblockA{\IEEEauthorrefmark{1}Department of Signal Processing \\ Tampere University of Technology, Tampere, Finland}
\IEEEauthorblockA{\IEEEauthorrefmark{2}Electrical Engineering Department \\ College of Engineering, Qatar University, Qatar}
\IEEEauthorblockA{\IEEEauthorrefmark{3} Corporate Sector Research  \& Advanced Engineering \\ Robert Bosch GmbH, Renningen, Germany}}


\maketitle

\begin{abstract}
Devices equipped with accelerometer sensors such as today's mobile devices can make use of motion to exchange information. A typical example for shared motion is shaking of two devices which are held together in one hand. Deriving a shared secret (key) from shared motion, e.g. for device pairing, is an obvious application for this. Only the keys need to be exchanged between the peers and neither the motion data nor the features extracted from it. This makes the pairing fast and easy. For this, each device generates an information signal (key) independently of each other and, in order to pair, they should be identical. The key is essentially derived by quantizing certain well discriminative features extracted from the accelerometer data after an implicit synchronization. In this paper, we aim at finding a small set of effective features which enable a significantly simpler quantization procedure than the prior art. Our tentative results with authentic accelerometer data show that this is possible with a competent accuracy ($76$\%) and key strength (entropy approximately $15$ bits). 
\end{abstract}



% \fi
%\IEEEpeerreviewmaketitle

\section{INTRODUCTION}
\label{sec:intro}
In mobile computing world, the notion of device pairing holds a great potential for short-term interactions, for example file transfer and payment, and long-term interactions such as device pairing with an accessory.  The most common approach to address device pairing issue is typing a PIN code or password into the involved devices \cite{pin}. However, this approach is impractical and as a result brings an overhead when there are many short-lived pairings. 

\begin{figure}[!htbp]
\begin{center}
\includegraphics[width=0.5\textwidth]{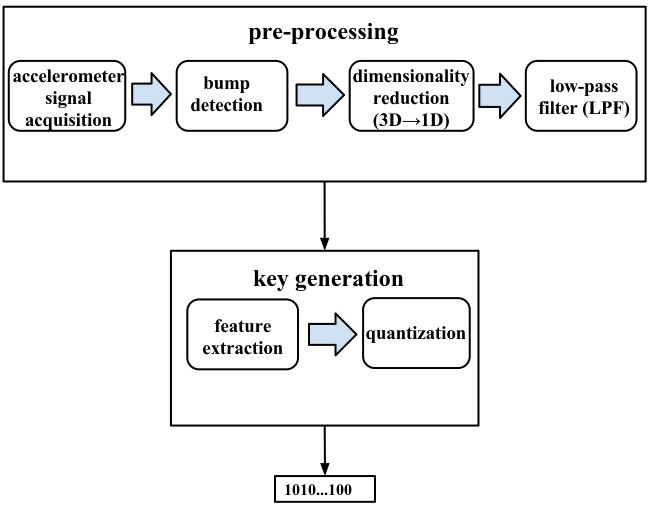}
\caption{The proposed approach}
\label{fig:architecture}
\end{center}
\end{figure}

Device pairing by shaking is a recent approach. It is a movement limited data channel between the two devices \cite{MayrhoferGellersen2009},\cite{ChongGellersen2012}. Shaking process consists of fast up and down movements in the $3$D space \cite{bichler}. It is known that two devices that are shaken together will experience similar but not exactly the same movement patterns \cite{shake}. This is a consequence of both imprecise accelerometer sensor embedded in the devices and different coordinate spaces of accelerometers during shaking process. In this work, this shared motion is exploited in that the recorded accelerometer signals are used to generate information signals independently of each other on both devices by feature extraction followed by a simple quantization procedure. Ideally, both information signals are expected to become identical. The proposed approach is shown in Figure~\ref{fig:architecture}.

In the proposed approach, any kind of communication between devices such as exchange of acceleration signal characteristics is not allowed until the confirmation process is accomplished. This means, for example, that the choice of using correlation of two signals was discarded. The ultimate goal is to generate an information signal from each accelerometer signal separately such that both information signals become identical for the same (shared) shaking processes. Moreover, both information signals should be different for different shaking processes. As a result of this, the problem of verification of whether the two devices are shaken together or not turns into a $2$-class classification problem.  For this particular aim, we first extract a few well-discriminative features such as kurtosis, crest factor, peak to peak, average power, etc. and then simply pass the feature signals to the standard decimal-to-binary quantizer equipped with a harsh rounding to obtain a binary key of a desired length
(key strength).    
 
The paper is organized as follows: In Section~\ref{related_work}, the related work is presented. The signal processing methodology is introduced in Section~\ref{methodology}. In Section~\ref{sec:experiments}, the experimental setup and results are presented and in Section~\ref{sec:conclusion} we conclude our work and discuss topics for future research.

\section{RELATED WORK}
\label{related_work}
Device pairing by shaking was first presented in ``Smart-Its Friends'' \cite{smf}. The drawback of this technique was lacking of authentication of the involved devices in the interaction. Another closely related work included accelerometer based analysis to determine whether the devices are carried by the same person or not \cite{lester}. The feature extraction algorithms used in \cite{shake} includes coherence measure, which was originally introduced in \cite{lester}, and quantized FFT coefficients. In \cite{bichler}, a similar approach in which acceleration signal was used for key generation was presented with a difference in using time domain acceleration features contrary to \cite{shake} where only frequency domain features were used. They claim that the same key would rather likely be generated for different shaking processes if the key generation was based on a frequency-based technique. They concluded that the frequency domain is not suitable for key generation.

Our work shows a similar approach proposed in \cite{bichler} and \cite{shake} which are, to our knowledge, the only two relevant articles\footnote{There exist several publications by the same authors based on practically the same ideas.}. The similarities between our work and \cite{bichler}, \cite{shake} lay on space dimension reduction using Euclidean norm, and both \cite{bichler} as well as Protocol $2$ in \cite{shake} prohibit exchange of acceleration
data. However, in \cite{bichler} the signal is divided into segments from which a few principal
components are extracted and used to learn the representation vectors. The key generation
algorithm in \cite{bichler} is based on pair wise nearest neighbour quantization. In contrast, we
pick a few features with high discrimination power. Thus, it is possible to use a computationally cheaper quantization method, namely standard decimal-to-binary quantization. In \cite{shake}, two alternative protocols are proposed. Protocol $1$ requires exchange of acceleration data which we prohibit. Protocol $2$ determines multiple candidate feature
vectors where one device transmits all candidate feature vectors to the other device.
Authentication is performed by thresholding the percentage of matching candidate
feature vectors. Contrary to \cite{shake}, we transmit only one information signal to each device which prevents the communication overhead. Another difference is that in \cite{shake} a protocol is implicitly triggered whereas we use explicit triggering by bumping both devices at the same time.

\section{METHODOLOGY}
\label{methodology}
\subsection{Pre-processing of accelerometer signals}
%\subsection{Pre-processing: Synchronization of accelerometer signals}
\label{sec:sync}
The first task is to acquire acceleration data properly from the accelerometer sensor. Since both mobile phones are unsynchronized initially, they need a user interaction for synchronizing the starting points of recording the shaking process (temporal alignment). This task can be realized through a direct user input such as pressing a ``start synchronization'' button. However, this method is not user-friendly. In this work, in order to make accelerometer sensors start data recording at the same time, two mobile phones are tapped simultaneously at one's hand. This tapping (or bump) leads to a significantly  high amplitude in acceleration signal since the two mobile phones were shaken and then are suddenly blocked. The acceleration signal values are compared to a pre-defined threshold value and when the bump occurs they both exceed the threshold value. Accordingly, it starts recording data. Alternatively, the user can initially start shaking with a fast movement so that two devices are subjected to a high acceleration. This also behaves as a bump and enables the synchronization. If the bump is detected by two devices, it is guaranteed that the two devices are synchronized within a few samples.

An accelerometer signal value\footnote{After excluding the effect of gravity $g$ (acquired through \textit{linear\_acceleration} sensor in Android API).} at a fixed time point is typically $3$-dimensional, say $(s_x,s_y,s_z)$. Since the spatial alignment between devices is unknown, the $3$ dimensions recorded by the two devices will not be aligned \cite{shake}.  In this work, Euclidean norm is thought to be convenient to circumvent this problem because shaking processes often take place approximately on only one fixed axis and hence e.g. the Euclidean norm enables us to see shaking processes as 1D oscillations (Figure~\ref{fig:raw_filter_diff}) \cite{bichler}, \cite{shake}. From now on, we call these 1D signals as raw acceleration signals, say $\textbf{s}$, whose elements $s=\sqrt{{s_x}^2+{s_y}^2+{s_z}^2}$ correspond to one time point each. Now, for example, the measured shared motion training data is the set $\mathcal{D}_1 = \{(\textbf{s}_i^1(m),\textbf{s}_i^2(m))\},$ where the index $i$ refers to a test subject who shakes the devices $1$ and $2$ simultaneously at one hand, $\textbf{s}$ is the above defined raw acceleration signal, and $m$ refers to the $m$th shake. For example, $\textbf{s}_i^1(m)$ is the raw acceleration signal (time series) obtained from the $m$th shaking process by the $i$th subject measured by the device $1$.  

Finally, the raw signals are filtered with a 1D box filter, which is a lowpass FIR filter. The lowpass filter smooths the noisy signal. If the kernel size is increased more, it starts to extract the envelope of the signal (see Figure ~\ref{fig:raw_filter_diff}).

\subsection{Feature extraction from accelerometer signals}
\label{sec:Fex}
The raw acceleration signals are typically high dimensional with respect to the time dimension (i.e. each time series contains a high number of elements) being impractical to work with. Feature extraction is applied not only to reduce the dimension but also to define an efficient collection of features to discriminate between shared and different shaking processes with a good accuracy. A feature signal is directly used to generate a binary information signal. 

In this work, $10$ different features were used: number of peaks, root-mean-square (rms), mean, variance, skewness, kurtosis, crest factor, peak to peak, autocorrelation and average power. These features are extracted from the whole acceleration signal without doing any windowing. Since the ranges of feature values are quite different, feature values are normalized before the feature signal is passed to the quantizer.

\subsection{Key generation}
The ultimate objective is to generate exactly the same key from shared shaking processes independently without exchanging any acceleration signal content. Moreover, we want our algorithm to generate different keys on devices when they are not shaken together. Mathematically speaking, for positive (shared motion) class test data $\D_1$ we would like to have
$\Q(\F(\textbf{s}_i^1(m)),nb)=\Q(\F(\textbf{s}_i^2(m)),nb)$ where $\Q(.)$ denotes the quantizer, $\F(.)$ denotes the feature extractor and $nb$ denotes the number of bits. On the other hand, for negative (different motion) class test data, say $\D_2$, we would like to have $\Q(\F(\textbf{s}_i^k(m)),nb) \neq \Q(\F(\textbf{s}_j^l(n)),nb)$ ($\D_1$ and $\D_2$ are mutually exclusive). The main assumption in key generation is that cross correlation between signals from the same shared motion is high enough to independently generate exactly the same key. Similarly, the cross correlation of signals from different shared motion is assumed to be small not to generate the same key. This would make it possible to define $\F(.)$ and $\Q(.)$ as given above such that the positive and negative classes can be separated with a sufficiently high accuracy and key strength.

\begin{figure}[!htbp]
\begin{center}
\includegraphics[width=0.5\textwidth]{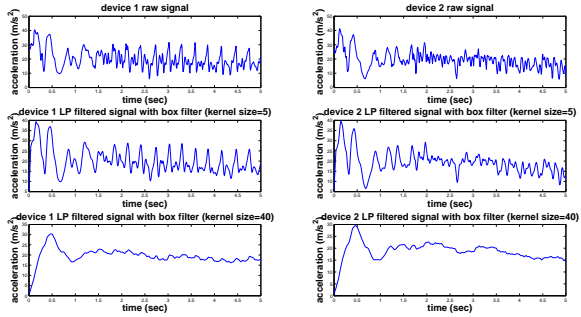}
\caption{Raw vs. filtered signals for device 1 and device 2}
\label{fig:raw_filter_diff}
\end{center}
\end{figure}

It is known that although the both signals are similar, they are not identical due to the reasons we discussed in Section~\ref{sec:intro}. For example, Figure~\ref{fig:raw_filter_diff} shows a similar but not identical signal pair. As a consequence, similar raw signals result in similar feature signals. However, we want our key generation algorithm to map similar feature signals to exactly same key which requires a hashing process. This could be realized via a quantizer which can also be interpreted as a classifier. 

Before the normalized feature signal is passed to $\Q(.)$, it is rescaled according to number of bits used in the binary representation of the key. The canonical conversion from decimal to binary is adopted for mapping. At the end of quantization and binary representation, a bit stream of a certain length will be generated based on the number of features and number of bits used in binary representation. 

The main task is to define $\F(.)$ and $\Q(.)$ using the test data $\D_1$ and $\D_2$ such that the classification accuracy between positive and negative classes is sufficiently high and, at the same time, a sufficiently strong key is possible to generate. This will be considered in detail in Section \ref{sec:experiments}. After defining $\F(.)$ and $\Q(.)$ the method is ready to be used for new shaking processes. When the two devices are shaken, each of them applies $\F(.)$ and $\Q(.)$ to their own acceleration signal respectively. It is important to notice that this procedure is autonomous which means that each device performs these computations locally without any knowledge about its peer. In the verification phase, device 1 transmits its information signal (bitstream) to device 2 and compares the received information signal with the result of its own and vice versa. If they are equal (either in strict sense or relaxed sense; c.f. Section \ref{sec:experiments}), the devices are paired. Otherwise, they are not paired which can be thought of as "access denied".

\section{EXPERIMENTAL SETUP AND RESULTS}
\label{sec:experiments}
In this work, two Samsung Galaxy Nexus smart-phones are used to acquire accelerometer sensor data. The data are acquired from \textit{linear\_acceleration} sensor, which is a software-based sensor, of Android API. The sampling rate $F_s$ of the sensor is $100$Hz. The off-line signal processing is performed using Matlab R2014a. 

The positive class test data $\D_1$ consists of $150$ shaking experiments recorded from 10 individuals ($i=1,\ldots,10;$ $m=1,\ldots,15$) . Five of the test subjects are male and five of them are female. All test subjects are asked to shake two devices ($1$ and $2$) together in one hand for five seconds which results in approximately $500$ time samples in an acceleration signal. Except this, no other instructions are given to the individuals. Negative class test data $\D_2$ is generated randomly from the positive class test data (i.e. by forming randomly pairs of $\textbf{s}_i^k(m)$ such that the pairs do not belong to $\D_1$). 
\begin{comment}
Test data:\\
$\omega_1$ : "positive class" ie. $\mathcal{D}_1 = \{(\textbf{s}_i^1(m),\textbf{s}_i^2(m))\}$ \\
$\omega_2$ : "negative class" ie. $\mathcal{D}_2 = \{(\textbf{s}_i^1(m),\textbf{s}_j^2(n))|\textbf{s}_i^1(m)\neq \textbf{s}_j^2(n)\} \setminus \mathcal{D}_1 $ \\
where the indices $i=1\dots10$ and $j=1\dots10$ refer to test subjects who shakes the two devices, $\textbf{s}$ is a row accelerometer signal, and $m=1\dots15$ and $n=1\dots15$ refer to test sample recordings (the $m$th shake and the $n$th shake). 
\end{comment}
The number of pairs of acceleration signals in $\mathcal{D}_1$ is $10\times15=150$. For $\D_2$, in turn, $300$ test samples are randomly generated from $\D_1$ such that first two random individuals are selected out of $10$ individuals and then two acceleration signals of those two individuals are randomly selected. This pair of signals constitutes one negative test sample of $\D_2$. 

To define $\F(.)$ we use the 10 features listed in Section \ref{sec:Fex}. We have chosen them manually in this work. The standard decimal-to-binary quantizer is applied to each of the 10 feature signals with the number $nb$ bits. The resulting 10 bitstreams are then concatenated resulting in keys of $10\times nb$ bits. This procedure forms the quantizer $\Q(.)$ in this article. It is worth to notice that this quantization method is very simple to implement and calculate. Now, we have to assess whether we can achieve a sufficiently good performance and keys strong enough with these definitions of $\F(.)$ and $\Q(.)$.

For performance assessment, confusion matrices\footnote{Let C be 2$\times$2 confusion matrix. According to matlab indexing C(1,1) is true positive (TP), C(1,2) is false negative (FN), C(2,1) is false positive (FP), C(2,2) is true negative (TN)} as well as accuracy\footnote{$Accuracy=\dfrac{TP+TN}{TP+TN+FP+FN}$} and $F1$\footnote{$F1=\dfrac{2TP}{2TP+FP+FN}$} measure are presented for both strict and relaxed cases (Table~\ref{tab:strict_best_param}, ~\ref{tab:relaxed_best_param}, ~\ref{tab:Confusion matrices}). In the strict case, both information signals must agree on every bit to be considered
a positive class whereas in the relaxed case, both bit strings must agree on at least $90$\% of all bits. For example, for a bit string of length $40$, at least $36$ bits are required to be same. Note that if both bit strings are close but not exactly the same, e.g. as in our case the Hamming distance is small, they can still be used to establish a secure encryption scheme based on a so-called fuzzy extractor \cite{dodis2008fuzzyextractor}.
\begin{comment}
The strict case refers to the case in which every bit of the information signals must be exactly same in order to be considered as a positive class whereas the relaxed case refers to the case where (at most) $10$\% of bit streams is allowed to be different. For example, for a bit stream of length $40$, at least $36$ bits are required to be same. Since at most $k$ bits out of $n$ bits (where $k$ is small)  are allowed to differ, it is still possible to construct encrypted keys with high probability and thus establish secure encryption based on so-called fuzzy extractors \cite{dodis2008fuzzyextractor}.
\end{comment}

\begin{figure}[!htbp]
\begin{center}
\includegraphics[width=0.5\textwidth]{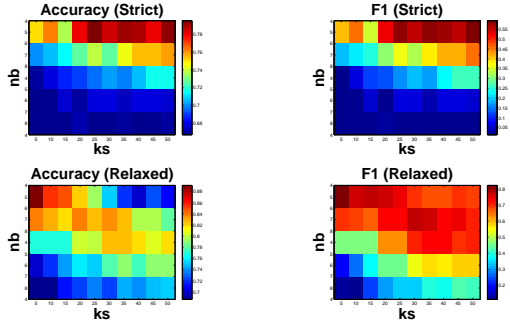}
\end{center}
\caption{Parameter space for filtered signals}
 \label{paramspace}
\end{figure}

The results presented in Table~\ref{tab:Confusion matrices} are obtained using the best values of \textit{nb} (refers to the number of bits in binary representation) and \textit{ks} (refers to the low pass filter size used in preprocessing stage). The best parameters of \textit{nb} and \textit{ks} in terms of maximizing accuracy and $F1$ are determined with an exhaustive grid search for both strict and relaxed case. 				Figure \ref{paramspace} shows parameter space for filtered signals.

\begin{table}[h]
\caption{Performance measure and best parameters in strict case (Acc=Accuracy, nb=number of bits, ks=kernel\_size)}
\centering
\begin{tabular}{l|l|l|l||l|l|l|}
\cline{2-7}
                            & Acc  & nb & ks & F1   & nb & ks \\ \hline
\multicolumn{1}{|l|}{RAW}   & 0.76 & 4  & -  & 0.46 & 4  & -  \\ \hline
\multicolumn{1}{|l|}{FILT.} & 0.79 & 4  & 25 & 0.59 & 4  & 50   \\ \hline
\end{tabular}
\label{tab:strict_best_param}
\end{table}

\begin{table}[h]
\caption{Performance measure and best parameters in relaxed case (Acc=Accuracy, nb=number of bits, ks=kernel\_size)}
\centering
\begin{tabular}{l|l|l|l||l|l|l|}
\cline{2-7}
                            & Acc  & nb & ks & F1   & nb & ks \\ \hline
\multicolumn{1}{|l|}{RAW}   & 0.84 & 4  & -  & 0.73 & 4  & -  \\ \hline
\multicolumn{1}{|l|}{FILT.} & 0.89 & 4  & 5 & 0.82 & 4  & 5   \\ \hline
\end{tabular}
\label{tab:relaxed_best_param}
\end{table}

% \usepackage{multirow}
\begin{table}[h]
\caption{Confusion matrices for all cases}
\begin{tabular}{c|cc|cc|cc|cc|}
\cline{2-9}
                                             & \multicolumn{4}{c|}{STRICT}                             & \multicolumn{4}{c|}{RELAXED}                            \\ \cline{2-9} 
                                             & \multicolumn{2}{c|}{Accuracy} & \multicolumn{2}{c|}{F1} & \multicolumn{2}{c|}{Accuracy} & \multicolumn{2}{c|}{F1} \\ \hline
\multicolumn{1}{|c|}{\multirow{2}{*}{RAW}}   & 46            & 104           & 46         & 104        & 99            & 51            & 99          & 51        \\ %\cline{2-9} 
\multicolumn{1}{|c|}{}                       & 0             & 300           & 0          & 300        & 21            & 279           & 21          & 279       \\ \hline
\multicolumn{1}{|c|}{\multirow{2}{*}{FILT.}} & 64            & 86            & 68         & 82         & 114           & 36            & 114         & 36        \\ %\cline{2-9} 
\multicolumn{1}{|c|}{}                       & 6             & 294           & 11         & 289        & 13            & 287           & 13          & 287       \\ \hline
\end{tabular}
\label{tab:Confusion matrices}
\end{table}

The confusion matrices as well as accuracy and $F1$ measures show that relaxing the key confirmation criteria obviously increases the performance. Lowpass filtering the raw acceleration signals has a positive effect on the results. The best accuracy and $F1$ results are gained with filtered signals in relaxed case using $4$ bits and a kernel size of $5$. As expected, the percentage of false negatives is higher than false positives.

For the above four cases (strict/relaxed, raw/filtered) we also estimated the entropies of the information signals. The maximal possible entropy is of course $40$ bits when each of the $10$ feature signals are quantized to $nb=4$ four bits and then concatenated to one bitstream of length $40$. The needed probabilities were obtained by estimating a multivariate Bernoulli mixture with the expectation maximization algorithm from our keys \cite{Juan}. The Bayesian information criterion was used to determine the size of the mixture \cite{Schwarz}. The hereby calculated entropies varied between $14$-$16$ bits for the four cases, which is sufficiently strong security for typical device pairing applications.

\section{CONCLUSIONS}
\label{sec:conclusion}
In this work, we have presented a recent idea of generating cryptographic key from the shared shaking movement using two smart phones. The generated key is going to be used for pairing of mobile phones which enables a secure connection between devices. The main idea in this work is that two devices shaken together in one hand experiences similar acceleration signals which can be utilized to generate a cryptographic key locally without any communication between devices until confirmation phase. 

In this paper, first we address the problem of synchronization in an efficient way. Then, we demonstrated that by utilizing only a few (10) informative features a strong 40-bit key could be generated. The average entropy was approximately $15$ bits per key which is slightly higher than the entropy of the Bluetooth PIN ($10$-$13$ bits). Off-line experiments showed that $76$\% of same shaking processes generate the same key with pre-processing and relaxing the key confirmation criteria. On the other hand, only $4$\% of different shaking processes generated the same key.

We conclude that our results ($76$\%, $4$\%) are promising and sufficiently accurate for our purposes. The proposed method allows generation of strong keys with a significantly simpler quantization method than in \cite{bichler}. The features were chosen here manually, but we are certain that it is possible to develop a sophisticated and objective feature extractor which can define the most optimal features based on a given training data in order to satisfy pre-defined accuracy and key strength requirements. This will be the topic of our future research.

\bibliographystyle{IEEEtran}
\bibliography{ShakeMe_PiComp2015}

\begin{thebibliography}{10}
\providecommand{\url}[1]{#1}
\csname url@samestyle\endcsname
\providecommand{\newblock}{\relax}
\providecommand{\bibinfo}[2]{#2}
\providecommand{\BIBentrySTDinterwordspacing}{\spaceskip=0pt\relax}
\providecommand{\BIBentryALTinterwordstretchfactor}{4}
\providecommand{\BIBentryALTinterwordspacing}{\spaceskip=\fontdimen2\font plus
\BIBentryALTinterwordstretchfactor\fontdimen3\font minus
  \fontdimen4\font\relax}
\providecommand{\BIBforeignlanguage}[2]{{%
\expandafter\ifx\csname l@#1\endcsname\relax
\typeout{** WARNING: IEEEtran.bst: No hyphenation pattern has been}%
\typeout{** loaded for the language `#1'. Using the pattern for}%
\typeout{** the default language instead.}%
\else
\language=\csname l@#1\endcsname
\fi
#2}}
\providecommand{\BIBdecl}{\relax}
\BIBdecl

\bibitem{pin}
C.~Gehrmann, C.~J. Mitchell, and K.~Nyberg, ``Manual authentication for
  wireless devices,'' \emph{RSA Cryptobytes}, vol.~7, no.~1, pp. 29--37, Spring
  2004.

\bibitem{MayrhoferGellersen2009}
R.~Mayrhofer and H.~Gellersen, ``Shake well before use: Intuitive and secure
  pairing of mobile devices,'' \emph{IEEE Transactions on Mobile Computing},
  vol.~8, no.~6, pp. 792--806, 2009.

\bibitem{ChongGellersen2012}
\BIBentryALTinterwordspacing
M.~K. Chong and H.~Gellersen, ``Usability classification for spontaneous device
  association,'' \emph{Personal Ubiquitous Comput.}, vol.~16, no.~1, pp.
  77--89, Jan. 2012. [Online]. Available:
  \url{http://dx.doi.org/10.1007/s00779-011-0421-1}
\BIBentrySTDinterwordspacing

\bibitem{bichler}
\BIBentryALTinterwordspacing
D.~Bichler, G.~Stromberg, M.~Huemer, and M.~L\"{o}w, ``Key generation based on
  acceleration data of shaking processes,'' in \emph{Proceedings of the 9th
  international conference on Ubiquitous computing}, ser. UbiComp '07.\hskip
  1em plus 0.5em minus 0.4em\relax Berlin, Heidelberg: Springer-Verlag, 2007,
  pp. 304--317. [Online]. Available:
  \url{http://dl.acm.org/citation.cfm?id=1771592.1771610}
\BIBentrySTDinterwordspacing

\bibitem{shake}
R.~Mayrhofer and H.~Gellersen, ``Shake well before use: Authentication based on
  accelerometer data,'' in \emph{In Pervasive}.\hskip 1em plus 0.5em minus
  0.4em\relax Springer, 2007, pp. 144--161.

\bibitem{smf}
L.~E. Holmquist, F.~Mattern, B.~Schiele, P.~Alahuhta, M.~Beigl, and H.-W.
  Gellersen, ``Smart-its friends: A technique for users to easily establish
  connections between smart artefacts,'' in \emph{Proc. Ubicomp 2001}, ser.
  LNCS, no. 2201, Springer-Verlag, 2001, pp. 116--122.

\bibitem{lester}
J.~Lester, B.~Hannaford, and G.~Borriello, ``'are you with me?' - using
  accelerometers to determine if two devices are carried by the same person,''
  in \emph{Proceedings of the Second International Conference on Pervasive
  Computing}, Vienna, Austria, 2004, pp. 33--50.

\bibitem{dodis2008fuzzyextractor}
Y.~Dodis, R.~Ostrovsky, L.~Reyzin, and A.~Smith, ``Fuzzy extractors: How to
  generate strong keys from biometrics and other noisy data,'' \emph{SIAM
  journal on computing}, vol.~38, no.~1, pp. 97--139, 2008.

\bibitem{Juan}
A.~Juan, J.~Garc{\'{\i}}a{-}Hern{\'{a}}ndez, and E.~Vidal, ``{EM}
  initialisation for bernoulli mixture learning,'' in \emph{Structural,
  Syntactic, and Statistical Pattern Recognition, Joint {IAPR} International
  Workshops, {SSPR} 2004 and {SPR} 2004, Lisbon, Portugal, August 18-20, 2004
  Proceedings}, 2004, pp. 635--643.

\bibitem{Schwarz}
G.~Schwarz, ``Estimating the dimension of a model,'' \emph{Ann. Statist.},
  vol.~6, no.~2, pp. 461--464, 03 1978.

\end{thebibliography}

\end{document}